\documentclass{elsart}

\usepackage{epsfig}

\newcommand{\Ket}[1]{\left\vert#1\right\rangle}

\begin{document}

\begin{frontmatter}



\title{A New Monomeric Interpretation of Intrinsic
Optical Bistability Observed in $Yb^{3+}$ - Doped Bromide
Materials}


\author[1]{F. Ciccarello}

\author[1]{A. Napoli}

\author[1]{A. Messina}
\ead{messina@fisica.unipa.it}

\author[2]{S.R. L\"{u}thi}

\address[1]{INFM and MIUR, Dipartimento di Scienze Fisiche ed
Astronomiche dell'Universit\`{a} di Palermo, via Archirafi 36,
90123 Palermo, Italy}

\address[2]{Gemfire Corporation, 1220 Page
Avenue, Fremont, CA - 94538, USA}

\begin{abstract}
We present a mechanism able to show intrinsic bistable behaviour
involving single $Yb^{3+}$ ions embedded into bromide lattices, in
which intrinsic optical bistability (IOB) has been observed. The
mechanism is based on the experimentally found coupling between
the $Yb^{3+}$ ion and the totally symmetric local mode of
vibration of the $[YbBr_{6}]^{3-}$ coordination unit. The model
reproduces the IOB observed in $CsCdBr_{3}:1\; \% \;Yb^{3+}$ and
allows to understand the experimentally found presence of the
phenomenon in the other bromides, but its absence in
$Cs_{3}Lu_{2}Cl_{9}:Yb^{3+}.$
\end{abstract}

\begin{keyword}
IOB \sep $Yb^{3+}$ \sep electron-phonon interaction
\PACS 42.65.Pc \sep 63.20.Kr \sep 63.20.Mt
\end{keyword}

\end{frontmatter}

A physical system without mirrors of an external cavity,
exhibiting two stable outputs for a single input intensity of an
exciting laser, is said to exhibit intrinsic optical bistability
(IOB) \cite{Abraham2,BowdenSung,Hopf}. Hehlen et al.
\cite{PRL94,JChemPhys96,LuthiJLumin,PRL99} have discovered that
$Yb^{3+}$ - doped $Cs_{3}X_{2}Br_{9}$ [$X = Y, Lu$] and
$CsCdBr_{3}$ show IOB when exciting into the
$^{2}F_{7/2}(0)\rightarrow ^{2}F_{5/2}(0')$ and
$^{2}F_{7/2}(0)\rightarrow ^{2}F_{5/2}(2')$ electronic transitions
of the $Yb^{3+}$ ion with a narrow band laser of suitable
near-infrared (NIR) frequency. On the contrary, no IOB was
observed under excitation into the $^{2}F_{7/2}(0)\rightarrow
^{2}F_{5/2}(1')$ transition \cite{TesiLuthi,Gamelin}. When doped
in these crystalline bromide materials some $Yb^{3+}$ ions form
strongly interacting $Yb^{3+}-Yb^{3+}$ dimers \cite{3cm-1,Mehta}.
Due to this interaction, excitation of the $Yb^{3+}$
$^{2}F_{7/2}\rightarrow ^{2}F_{5/2}$ transition (around $10000 \;
cm^{-1}$) results in both, $^{2}F_{5/2}\rightarrow ^{2}F_{7/2}$
emission of single $Yb^{3+}$ ions in the NIR spectral range, as
well as $Yb^{3+} - Yb^{3+}$ dimer emission in the visible (VIS) at
twice the resonance frequency of the single $Yb^{3+}$ ion
\cite{LuthiJLumin,1768,AuzelJLumin} through a cooperative
luminescence mechanism \cite{Auzel}. The IOB phenomenon consists
of a temperature and excitation power dependent hysteretic
behaviour of the transmitted, NIR and upconverted VIS emission
intensity (with the system switching output intensity at two
critical incident excitation powers). To date, it is discussed if
the phenomenon is due to the presence of strongly interacting
$Yb^{3+}-Yb^{3+}$ dimers or if it is a single $Yb^{3+}$ ion
property. The theories relying on the former interpretation are
usually referred to as \lq\lq dimeric\rq\rq while those supporting
the latter idea as \lq\lq monomeric\rq\rq. Dimeric theories
substantially rely on the existence of intra-pair electronic
coupling of $Yb^{3+}-Yb^{3+}$ dimers
\cite{PRL94,JChemPhys96,GuillotPRB,GuillotPRA}. According to these
theories, IOB should be observed \emph{a fortiori} in $Yb^{3+}$ -
doped $Cs_{3}Lu_{2}Cl_{9}$, as this material is isostructural to
$Yb^{3+}$ - doped $Cs_{3}X_{2}Br_{9}$ [$X = Y, Lu$]
\cite{Meyer,Schleid} and exhibits a stronger electronic exchange
interaction for $Yb^{3+}-Yb^{3+}$ dimers \cite{3cm-1}. However, no
experimental evidence for IOB was found in $Yb^{3+}$-doped
$Cs_{3}Lu_{2}Cl_{9}$ \cite{TesiLuthi,Gamelin}. Furthermore, in
$Cs_{3}X_{2}Br_{9}:Yb^{3+}$ ($X=Y,Lu$) only a few $Yb^{3+}$ ions
form dimers at low concentration of impurity ions. Nevertheless,
both bromide lattices exhibit IOB even at a low $Yb^{3+}$ - doping
level \cite{PRL94,JChemPhys96,LuthiJLumin}. A monomeric mechanism
to account for IOB in these systems has been proposed by Gamelin
et al. \cite{Gamelin} who presented a phenomenological model,
according to which IOB depends on a single - $Yb^{3+}$ ion
property in these materials. In this model, IOB is attributed to
the interplay of a laser heating effect and a strongly increasing,
non-linear dependence of the material's absorbance with the
internal temperature. With its phenomenological nature, this
theory succeeds in reproducing many experimental trends, but does
not provide the microscopic mechanism of IOB. In this letter, we
present the first microscopic, monomeric mechanism that allows a
single active ion to exhibit all essential aspects observed in
IOB. To date, the microscopic theories presented in literature
describing how a collection of two-level atoms can exhibit IOB are
based on electronic coupling between two or more near absorbers
\cite{Hopf,JChemPhys96,GuillotPRB,Heber87}. In these cases, the
electronic interaction between the coupled absorbers provides the
feedback loop required to obtain intrinsic bistability. Together
with the intrinsic two-level atom non linearity and the use of
single-atom density matrix approach (an indispensable step in
order to obtain non linear equations), this feedback loop gives
rise to inversion dependent atomic resonance frequency
renormalization, and thus possible IOB. With a few exceptions, the
role played by the host lattice, and in particular of the first
coordination sphere of the $Yb^{3+}$, in the mechanism responsible
for IOB in $Yb^{3+}$-doped bromide host materials has generally
been regarded as secondary. However, Hehlen  et al. measured
different internal temperatures for the system on the two branches
of the hysteresis loop \cite{JChemPhys96}. This suggests that with
the first switch occurring, the lattice undergoes heating.
Additionally, L\"{u}thi \cite{TesiLuthi} has found that
$Cs_{3}Lu_{2}Br_{9}:10  \% \; Yb^{3+}$ shows bistable behaviour
even when excited with a chopped laser and off-times as long as
$75\;msec$. This means that the system prepared on the upper
branch of the bistability region does not decay to the lower
branch for times as long as $75\;msec$ after turning off the
exciting laser beam. If IOB were due to a cooperative mechanism
based on electronic $Yb^{3+}-Yb^{3+}$ coupling, and the feedback
loop provided by the same two ytterbium ions, it would be hard to
justify this behavior. After $75\;msec$, all ions in their excited
state are expected to have decayed to the ground state, and there
seems to be no physical reason for the system to save memory of
its past history. The fact that the nuclear relaxation times of
the lattice are much longer than the electronic ones, suggests
that the lattice nuclear degrees of freedom play a relevant role
in the bistability mechanism. Supported by the findings of a Raman
spectroscopy study on $CsCdBr_{3}:1\%\;Yb^{3+}$
\cite{electron-phonon} showing strong coupling between the
electronic $^{2}F_{7/2}(0)\rightarrow ^{2}F_{5/2}(2')$ transition
of $Yb^{3+}$ and the totally symmetric vibrational local mode of
the $[YbBr_{6}]^{3-}$ coordination unit (into which $Yb^{3+}$ is
embedded in all bromide lattices showing IOB), we propose that
this vibrational coupling provides the indispensable feedback loop
to observe IOB (involving only a single $Yb^{3+}$ ion).

Our physical system consists of the electronic structure of a
single $Yb^{3+}$ ion and the totally symmetric vibrational local
mode $A_{1g}$ of the $[YbBr_{6}]^{3-}$ octahedron. We treat the
former as a two-level system ($\Ket{a}$ and $\Ket{b}$ ground and
excited states, respectively, $\hbar \omega_{0}$ energy
difference, $\Ket{i}\langle j |=\sigma_{ij}$ $i,j=a,b$) and the
latter as a quantum harmonic oscillator ($a$ and $a^{+}$
annihilation and creation operators, respectively, and angular
frequency $\omega_{k}$). We write the free Hamiltonian $H_{0}$ of
the system as
\begin{equation} \label{Ho}
H_{0}=H_{Yb}+ H_{k}= \hbar \omega _{0} \; \sigma_{bb}+\hbar
\omega_{k} \; a^{+}a
\end{equation}
For the interaction Hamiltonian $V_{Yb-k}$ between the ion and the
$A_{1g}$ local mode we choose a bilinear operatorial form (in
order for the oscillation centre to depend on the electronic state
in which the ion is \cite{Pryce})
\begin{equation} \label{VYb-k}
V_{Yb-k}=\epsilon \; (\sigma_{bb}-\sigma_{aa})\;(a+a^{+})
\end{equation}
When a coherent classical electromagnetic field is applied to the
ion, using the rotating wave and the electric dipole
approximations, the coupling between the $Yb^{3+}$ ion and the
external field is described by the observable
$V_{Yb-L}=\frac{\hbar \Omega}{2}(\sigma_{ba} e^{-i \omega t}+
\sigma_{ab} e^{i \omega t})$, where $\Omega$ and $\omega$ stand
for the Rabi and the laser frequency. Denoting the total
Hamiltonian operator of the system with $H = H_{0}+ V_{Yb-k} +
V_{Yb-L}$, and taking into account both the ion spontaneous
emission and the dissipative interaction of the ion and the local
mode with the bulk modes of the lattice (described by a suitable
superoperator $S$), the dynamics of the system obeys the
generalized Von-Neumann equation
\begin{equation} \label{equazioneF}
\dot{F}= - \frac{i}{\hbar}\left[ H,F \right]+SF
\end{equation}
where $F$ is the density operator of the whole system. $F(t)$ can
be written as $F(t)=Tr_{k}F(t) \; Tr_{Yb}F(t)+ F^{Yb-k}(t)$, where
$Tr_{k}F$ and $Tr_{Yb}F$ are the reduced density operators of the
ion and the local mode, respectively, while $F^{Yb-k}(t)$
represents the correlations existing at the time $t$ between $Yb$
and $k$. Assuming $V_{Yb-k}\ll H_{0}$, we approximate $F(t)$ with
its factorized part
\begin{equation} \label{HpFatt}
F(t)\simeq Tr_{k}F(t) \; Tr_{Yb}F(t)
\end{equation}
This is the main hypothesis of our model. In this way, analogously
to other theories \cite{Hopf,JChemPhys96,Heber87,Bodenshatz}, we
\emph{assume} that single-ion density matrix approach is allowed.
A necessary condition for this approach to be valid is that
$V_{Yb-k}$ is much smaller than the free Hamiltonian $H_{0}$
(later on we will show this is the case). It must be remarked that
the feasibility of this treatment - indispensable to have non
linear equations and thus possible IOB - poses a delicate and
lively debated \cite{GuillotPRB,Malyshev,Heber2000} theoretical
problem questioning even the existence of a microscopic
explanation of some IOB phenomena \cite{Malyshev}.  We justify
assumption (\ref{HpFatt}) with the ability of our model to
reproduce the observed IOB phenomena. Inserting eq.(\ref{HpFatt})
in eq.(\ref{equazioneF}) and doing the trace over $k$ and $Yb$, we
obtain two coupled master equations in $Tr_{k}F(t)$ and
$Tr_{Yb}F(t)$. Performing the variable change
$\rho'_{ab}=\rho_{ab} \; e^{i \omega t}$, it can be proved that
the ion dynamic equations in $\dot{\rho}_{bb}$ and
$\dot{\rho'}_{ab}$ differ from the usual optical Bloch equations
in the rotating frame for the presence of an effective resonance
frequency
$\omega_{0}^{eff}(t)=\omega_{0}+2\frac{\epsilon}{\hbar}\langle a +
a^{+}\rangle (t)$, linearly dependent on the mean value of the
local mode operator $\left( a+a^{+} \right)$. This proves that the
dynamic response of the ion to the applied field depends on the
state of the local mode (in its turn depending on the ion state).
It is easy to show that for stationary conditions (when setting
all temporal derivatives to zero) the resonance frequency can be
written as
\begin{equation} \label{frequenzaeffettivaregime}
\omega_{0}^{eff}= \omega_{0}- \beta (\rho_{bb}-\rho_{aa})
\end{equation}
where
\begin{equation} \label{beta}
\beta = 4 \; \frac {\epsilon^{2}}{\hbar^{2}} \; \frac
{\omega_{k}}{\omega_{k}^{2}+ \gamma^{2}}
\end{equation}
with $\gamma$ being the relaxation rate of the local mode. We thus
find that the overall effect of the coupling between the
electronic transition of the $Yb^{3+}$ ion and the $A_{1g}$ local
mode of the $[YbBr_{6}]^{3-}$ unit, consists of an ion resonance
frequency renormalization linearly dependent on inversion
($\rho_{bb}-\rho_{aa}$). Introducing the absorption parameter $s=2
\rho_{bb}$ ($0\le s\le 1$), it is possible to prove that, under
stationary conditions, $s$ and the incident excitation power $P$
must obey the equation
\begin{equation} \label{eqcubicasgammah}
c_{3} \; s^{3}+ c_{2} \; s^{2}+ \left( c_{1} + b \; P\right) \; s
- b \; P=0
\end{equation}
with
\begin{eqnarray} \label{coefficienti}
\nonumber c_{3}= 4 \; \left(\frac{\beta}{\Gamma_{h}} \right)^{2}
\; c_{2}= -8 \; \frac{\beta}{\Gamma_{h}}  \;  \frac{\delta +
\beta}{\Gamma_{h}}\;\;c_{1}= 4 \; \left( \frac{\delta + \beta}
{\Gamma_{h}} \right)^{2}+1 \\
b = \frac{3e^{2}}{\pi mc\epsilon_{0}}\;  \frac{F_{ab}}{\hbar
\omega_{0}} \;cos^{2}\varphi \; \frac {1}{\Gamma_{1} \Gamma_{h}}
\;\;\;\;\;\;\;\;\;\;\;\;\;\;\;\;\;\;\;\;\;\;\;\;
\end{eqnarray}
where $\Gamma_{h}$ is the homogeneous linewidth of the electronic
transition, $\delta = \omega_{0}- \omega $ is the detuning of the
laser frequency, $F_{ab}=(2m\omega_{0}/3\hbar)|\langle
a|\mathbf{r}|b\rangle|^{2}$ stands for the oscillator strength of
the transition $\Ket{a}\rightarrow\Ket{b}$, $\varphi$ is the angle
between the laser polarization versor and $\langle
a|\mathbf{r}|b\rangle$, and $\Gamma_{1}$ is the ion longitudinal
decay time. Eq.(\ref{eqcubicasgammah}) is a third degree equation
in the unknown $s$ dependent on many parameters. It can be shown
that the condition for the existence of a $P$-range inside which,
for any value of $P$, two stable solutions for $s$ are found (IOB)
is
\begin{equation} \label{condizioneIOB}
\frac{\beta}{\Gamma_{h}}\ge \frac{3\sqrt{3}}{2}
\end{equation}
This provides a threshold value of the local mode - $Yb^{3+}$ ion
interaction strength required to observe intrinsic bistability
compatibly with our microscopic feedback mechanism. Moreover,
eq.(\ref{eqcubicasgammah}) shows that, for an increasing value of
the bistability parameter $\beta/\Gamma_{h}$, the incident power
bistability range moves toward higher values of $P$ and, in
addition, becomes wider. We now use our theory to reproduce the
two incident powers at which switching occurs measured in
$CsCdBr_{3}:1\;\% \; Yb^{3+}$ \cite{PRL99}. Guillot-No\"{e}l et
al. successfully performed an analogous simulation to test their
dimeric theory \cite{GuillotPRB}. In our study we use: $\delta=0$
(perfect resonance), $\omega_{0}=10602.8 \; cm^{-1}$ \cite{PRL99},
$\Gamma_{h}=0.6 \; cm^{-1}$ \cite{PRL99}, $1/\Gamma_{1}=7.8 \cdot
10^{-4} s$ \cite{AuzelJLumin}, $F_{ab}= 10^{-6}$ and,
additionally, we take $cos^{2}\varphi=1/2$, assuming that the
incident laser beam is unpolarized \cite{JChemPhys96}. Using the
bistability parameter to optimise our simulation, good agreement
is achieved between the theoretical prediction and the
experimentally measured values \cite{PRL99} for the two incident
powers at which switching occurs, as shown in Fig. 1.
 \begin{figure}
\includegraphics{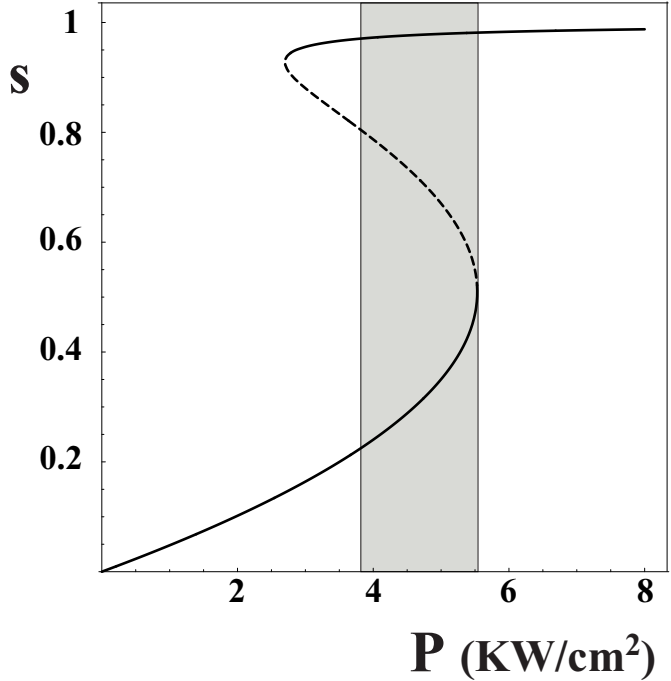}
\caption{Absorption parameter $s$ of the $Yb^{3+}$ ion as a
function of the incident intensity $P$ ($kW/cm^{2}$) in
$CsCdBr_{3}:1\% \; Yb^{3+}$ at $7 \; K$ calculated from eq.
(\ref{eqcubicasgammah}) ($\beta/\Gamma_{h}=7.5$, $\delta=0$,
$\omega_{0}=10602.8 \; cm^{-1}$, $\Gamma_{h}=0.6 \; cm^{-1}$,
$1/\Gamma_{1}=7.8 \cdot 10^{-4} sec$, $F_{ab}=10^{-6}$). The
dashed line represents unstable solutions. The grey shaded area
represents the experimentally observed bistable region for this
compound \cite{PRL99}.}
\end{figure}
Using the optimised value of $\beta/\Gamma_{h}=7.5$, the ion-local
mode coupling constant $\epsilon$ [eq.(\ref{VYb-k})] can be
estimated. From eq.(\ref{beta}) and taking into account the fact
that the lattice nuclear relaxation times are typically very long
compared to $1/\omega_{k}$, we get $\epsilon \simeq
\frac{\hbar}{2}\sqrt{\beta \; \omega_{k}}$. With $\omega_{k}=191
\;cm^{-1}$ \cite{electron-phonon}, we obtain $\epsilon=14.66 \;
cm^{-1}$ for $CsCdBr_{3}:1\%  \; Yb^{3+}$. This value confirms our
\emph{a priori} assumption that $V_{Yb-k}$ is much weaker than
$H_{0}$. With this value for $\epsilon$, the displacement of the
oscillation centre for the local mode $k$, for an $Yb^{3+}$ ion
undergoing the transition from $\Ket{a}$ to $\Ket{b}$, as implied
by the operatorial form assumed for $V_{Yb-k}$ [eq.(\ref{VYb-k})],
turns out to be in the order of $1\;\%$ . This result agrees with
typical values found for lattice point defects due to
incorporation of rare earth impurity ions \cite{Pryce}, and
confirms that the coupling constant $\epsilon$ we get from our
model is of the correct order of magnitude.

Another relevant test for our theory is its ability to predict the
presence of IOB in $Cs_{3}X_{2}Br_{9}: Yb^{3+}$ [$X=Y,Lu$], but
the absence of the phenomenon in $Cs_{3}Lu_{2}Cl_{9}: Yb^{3+}$
\cite{TesiLuthi,Gamelin}. In the bromide compounds, the angular
frequency $\omega_{k}$ is $190\;cm^{-1}$ \cite{1768}, similar to
the one found in $Cs_{3}Er_{2}Br_{9}$ \cite{HehlenPRB}. From
comparison of the absorption and emission spectra, we estimate
$\Gamma_{h}$ and $\epsilon$ to be comparable to the ones measured
in $CsCdBr_{3}:Yb^{3+}$ \cite{TesiLuthi,electron-phonon}. These
assumptions made, we expect the bistability parameter
$\beta/\Gamma_{h}$ for $Cs_{3}X_{2}Br_{9}: Yb^{3+}$ [$X=Y,Lu$] to
have a comparable value to the one calculated for $CsCdBr_{3}:1\%
\; Yb^{3+}$, and can understand why IOB is found in the former
compounds. For the chloride, $\omega_{k}$ is $285\;cm^{-1}$
\cite{HehlenPRB}. From inspection of the absorption and emission
spectra, we estimate $\Gamma_{h}$ to fall in the range
$0.3-0.7\;cm^{-1}$ \cite{TesiLuthi}, and the $Yb^{3+}$ - $A_{1g}$
local mode coupling constant $\epsilon$ to be about 5 times
smaller than in bromide compounds. Based on these values, the
bistability parameter $\beta/\Gamma_{h}$ is estimated to be lower
than the critical threshold of $3\sqrt{3}/2$ required for IOB. We
finally notice that absorption spectra of
$Cs_{3}Lu_{2}Br_{9}:Yb^{3+}$ show that the $Yb^{3+}$ ion
$^{2}F_{7/2}(0)\rightarrow ^{2}F_{5/2}(1')$ transition exhibits a
very weak coupling to the $A_{1g}$ local mode compared to $0-0'$
and $0-2'$. This explains why no IOB is observed under excitation
into this peculiar transition.

In summary, we have presented the first microscopic mechanism
according to which a single $Yb^{3+}$ ion is able to exhibit IOB
when doped in $Cs_{3}X_{2}Br_{9}$ [$X=Y, Lu$] or $CsCdBr_{3}$. The
mechanism couples the electronic two-level system of the $Yb^{3+}$
ion with the totally symmetric local mode of the vibration of the
$[YbBr_{6}]^{3-}$ coordination unit. The theory has been used to
reproduce basic aspects of the IOB observed in $CsCdBr_{3}:
1\%\;Yb^{3+}$. We emphasize that our theory not only allows to
understand, why IOB has been observed in $CsCdBr_{3}:
1\%\;Yb^{3+}$ and $Cs_{3}X_{2}Br_{9}$ [$X=Y, Lu$], but also, why
it is absent in the chloride compound $Cs_{3}Lu_{2}Cl_{9}:
Yb^{3+}$. These results provide new insight into the microscopic
mechanism underlying the IOB phenomenon and guidelines for
expanding the material basis in the search for IOB systems, and
their potential applications.

We gratefully acknowledge inspiring discussions with O.
Guillot-No\"{e}l of ENSCP, Paris and M.P. Hehlen of M. Lujan Jr.
of Neutron Scattering Centre, Los Alamos National Laboratory, Los
Alamos, New Mexico.

\emph{Note}: Octahedral coordination of the $[YbBr_{6}]^{3-}$
unit, instead of the actual $C_{3}$  symmetry, is assumed
throughout the text with no impact on our conclusions.

\end{document}